\documentclass[prl,aps,a4,twocolumn,superscriptaddress,preprintnumbers,nofootinbib]{revtex4-1}
\pdfoutput=1

\usepackage{pslatex}
\usepackage[pdftex]{graphicx}
\usepackage{psfrag}
\usepackage{epsfig}
\usepackage{color}
\usepackage{cancel}
\usepackage{slashed}
\usepackage{amssymb}
\usepackage{amsmath}
\usepackage{hyperref}
\usepackage{enumerate}
\usepackage{multirow}
\usepackage[normalem]{ulem}

\definecolor{vdrgreen}{rgb}{0.0, 0.6, 0.0}

\begin{document}

\title{Reactor neutrino background in third-generation dark matter detectors}%

\author{D. Aristizabal Sierra}%
\email{daristizabal@uliege.be}%
\affiliation{Universidad T\'ecnica Federico Santa Mar\'{i}a -
  Departamento de F\'{i}sica\\ Casilla 110-V, Avda. Espa\~na 1680,
  Valpara\'{i}so, Chile}%
\author{Valentina De Romeri}%
\email{deromeri@ific.uv.es}%
\affiliation{Instituto de F\'{\i}sica Corpuscular (CSIC-Universitat de
  Val\`encia), Parc Cient\'{\i}fic UV C/ Catedr\'atico Jos\'e
  Beltr\'an, 2 E-46980 Paterna (Valencia) - Spain}%
\author{Christoph A. Ternes}%
\email{christoph.ternes@lngs.infn.it }%
\affiliation{Istituto Nazionale di Fisica Nucleare (INFN), Laboratori
  Nazionali del Gran Sasso,\\ 67100 Assergi, L’Aquila (AQ), Italy}%
\begin{abstract}
  Third-generation dark matter detectors will be fully sensitive to
  the boron-8 solar neutrino flux. Because of this, the
  characterization of such a background has been the subject of
  extensive analyses over the last few years. In contrast,
  little is known about the impact of reactor neutrinos. In this
  letter we report on the implications of such a flux for dark matter
  direct detection searches. We consider five
  potential detector deployment sites envisioned by the recently established
  XLZD consortium: SURF, SNOLAB, Kamioka, LNGS and Boulby. By using
  public reactor data we construct five reactor clusters---involving
  about 100 currently operating commercial nuclear reactors each---and
  determine the net neutrino flux at each detector site. Assuming a
  xenon-based detector and a 50 tonne-year exposure, we show that in
  all cases the neutrino event rate may be sizable, depending on energy
  recoil thresholds. Of all possible detector sites, SURF and LNGS are
  those with the smallest reactor neutrino background.
  On the contrary, SNOLAB and Boulby are subject to the strongest reactor neutrino fluxes, with
  Kamioka being subject to a more moderate background. Our findings
  demonstrate that reactor neutrino fluxes should be taken into
  account in the next round of dark matter searches. We argue that
  this background may be particularly relevant for directional
  detectors, provided they meet the requirements we have employed in
  this analysis.
\end{abstract}
\maketitle
\section{Introduction}
\label{sec:intro}
A wealth of cosmological and astrophysical data supports the idea that
the dominant form of matter in the Universe has feeble or none
electromagnetic interactions. The \textit{conventional wisdom} is that
this new form of matter---dubbed dark matter (DM)---is of microscopical
origin and its abundance is determined by fast-scattering processes
with Standard Model (SM) particles at very early epochs, much
before the onset of cosmic neutrino decoupling and primordial
nucleosynthesis (for a review see
e.g. Ref. \cite{Arcadi:2017kky}). Although at high temperatures DM
is thermalized, as the temperature decreases---because of the expansion of the Universe
---these scattering processes are unable to keep the species
in thermodynamic equilibrium and so its abundance freezes out. This
weakly interacting massive particle (WIMP) is a rather generic
candidate appearing in a large class of particle physics models. It is a
dominant paradigm that has driven DM searches.

DM direct detection is a subject that dates back to the mid 80's, when
Goodman and Witten pointed out that WIMPs could be searched for by
using the same detectors proposed by Drukier and Stodolsky for
coherent elastic neutrino-nucleus scattering (CE$\nu$NS) measurements
\cite{Goodman:1984dc,Drukier:1984vhf}. Since then, and because of the
lack of a signal, detector technologies as well as fiducial volumes
have dramatically evolved. At present, DM searches in direct detection
experiments are led by liquid xenon (LXe) dual-phase time projection
chambers (second-generation DM detectors). Detectors at the INFN
``Laboratori Nazionali del Gran Sasso'' (LNGS) in Italy (XENONnT), at
the Sanford Underground Research Facility (SURF) in South Dakota in
the US (LZ) and at the China Jinping Underground Laboratory in
Sichuan, China (PandaX-4T) are using active volumes of the order of 5
tonne \cite{XENON:2023cxc,LZ:2022lsv,PandaX:2022xqx}.

With their high capabilities for background rejection, along with low
nuclear recoil energy thresholds, these second-generation DM detectors
are sensitive to spin-independent WIMP-nucleon total cross sections of
the order of $10^{-48}\,\text{cm}^2$ \cite{Billard:2021uyg}. Indeed, XENONnT and LZ have
recently published results where sensitivities of the order of
$\sigma_\text{WIMP-nuc}\sim 10^{-47}\,\text{cm}^2$ have been reported
\cite{XENON:2023cxc,LZ:2022lsv}. PandaX-4T has set the most
stringent upper limit in the low WIMP mass region ($\lesssim 10$ GeV),
$\sigma_\text{WIMP-nuc}\sim 10^{-44}\,\text{cm}^2$
\cite{PandaX:2022xqx}.

A new generation of LXe detectors---third-generation DM detectors---is
expected to pave the way for a discovery \footnote{Note that if a
  discovery takes place in second-generation detectors, the
  experimental environment provided by their third-generation follow-ups 
  will allow precise measurements of WIMP properties.}. Recently
the XENONnT, LZ and DARWIN collaborations have united forces and
created the XLZD consortium \cite{XLZD}. Their goal is the
construction of a 40-100 tonne detector with unprecedented
sensitivities. With such active volume, a detector of this kind will be subject to an irreducible neutrino background dominated
by $^8\text{B}$ solar neutrinos (for nuclear-channel signals) and by
$pp$ neutrinos (for electron-channel signals)~\cite{Strigari:2009bq}.

The morphology and size of this background have been the subject of
different analyses in recent years, first identified as the so-called
``neutrino floor''
\cite{Monroe:2007xp,Vergados:2008jp,Strigari:2009bq,Billard:2013qya,OHare:2016pjy}
and its more recent redefinition, the ``neutrino fog''
\cite{OHare:2021utq}, where a first estimation of the reactor neutrino
background at LNGS was addressed. It is well known that the impact of
the neutrino background on a WIMP discovery signal is mainly dominated
by neutrino flux uncertainties, with uncertainties on the weak mixing
angle and on the root-mean-square radii of the neutron distributions
playing a rather subdominant role
\cite{AristizabalSierra:2021kht}. The presence of a neutrino
background, however, does not mean that an identification of a WIMP
signal is impossible. First of all, improvements in the determination
of solar neutrino flux uncertainties are expected. Secondly, WIMP and
neutrino spectra in general do not fully degenerate in most regions of
parameter space. Even in regions where they strongly do, an
identification is possible with sufficiently large data sets
\cite{Billard:2013qya}. Furthermore, even if data is not abundant,
directionality will---potentially---enable a distinction between WIMP
and neutrino nuclear recoil spectra \cite{Vahsen:2021gnb}, if they
turn out to be strongly degenerate.

Given this landscape, and the fact that DM direct detection will soon
enter the third-generation detector phase, one should wonder whether
other neutrino sources might contribute to the background and hence
should be taken into account. This is a rather relevant question to
raise, aiming to leverage the full discovery power of these types of
detectors. Motivated by this question, in this Letter we assess the
impact of nuclear reactor neutrinos.  Since the reactor neutrino flux
strongly depends on the geographical position of the detector---for
definitiveness---we use LNGS, SURF, Boulby (UK), Kamioka (Japan) and
SNOLAB (Canada) as possible deployment sites \footnote{These
  underground facilities are considered as potential locations for
  detector deployment by the XLZD Consortium \cite{XLZD}.}.

\section{Nuclear reactor sources: Locations and event rates}
\label{sec:nuclear_reactor_sources}
The data sets we employ follow from data provided on the
 \href{https://reactors.geoneutrinos.org/}{Geoneutrinos.org} website~\cite{Dye:2015bsw,geoneutrinos}. We consider
only commercial power plants (that involve the most powerful
reactors) for which a non-zero operating power is reported. Reactors
for which the thermal capacity is known but have zero operating power
and those that have been permanently
shut down are not included. Depending on the baseline, each 
detector site that we consider is ``surrounded'' by a cluster of
nuclear reactor power
plants, at a certain distance $L_i$. Table~\ref{tab:minimum_and_maximum_baselines_power} shows
the minimum and maximum baseline and power for each cluster, along with
the number of reactors involved. For each detector site, we do not include reactors located at distances beyond $L_\textrm{max}$, as their contribution to the event rates would be negligible.

\begin{table}[!ht]
  \centering
  \begin{tabular}{|c||c|c|c|c|c|}\hline
    Location & NR &$L_\text{min}$ [km] & $L_\text{max}$ [km] & $P_\text{min}$ [GW]
    & $P_\text{max}$ [GW]\\\hline\hline
    SURF & 111 &790 & 2951& 0.34& 3.9\\\hline
    SNOLAB & 104 &239 & 2874& 0.92& 3.9\\\hline
    Kamioka & 86 &146 & 2895& 0.15& 3.9\\\hline
    LNGS & 146 &417 & 4027& 0.42& 3.7\\\hline
    Boulby & 141 &26 & 3654& 0.51& 3.7\\\hline
  \end{tabular}
  \caption{Minimum and maximum baselines ($L_\text{min}$ and
    $L_\text{min}$) along with minimum and maximum reactor powers
    ($P_\text{min}$ and $P_\text{max}$) for the SURF, SNOLAB, Kamioka,
    LNGS and Boulby reactor clusters. The number of
    reactors in each cluster (NR) is also shown. Data has been extracted from the
    \href{https://reactors.geoneutrinos.org/}{Geoneutrinos.org} website.}
  \label{tab:minimum_and_maximum_baselines_power}
\end{table}

\begin{figure*}[t]
  \centering
  \includegraphics[scale=0.45]{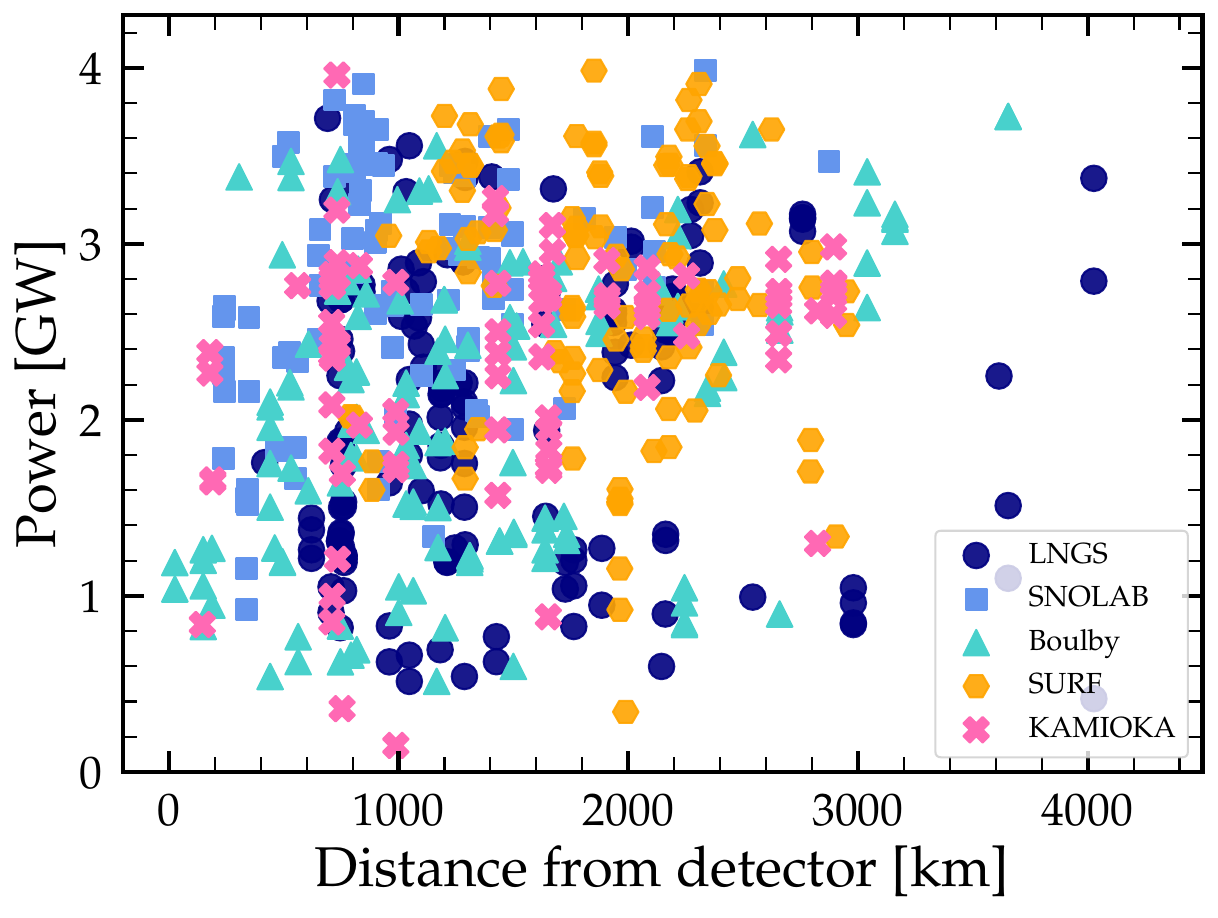}
  \caption{Location of the different reactors within the SURF, SNOLAB,
    Kamioka, LNGS and Boulby clusters and their corresponding
    operating power.}
  \label{fig:reactor_clusters}
\end{figure*}
The largest clusters are those around the LNGS and Boulby detector
sites (as expected, given that for these two cases the radius defining
the cluster exceeds by about 1000 km the radius at the other sites). 
However, this does not
necessarily mean that the largest flux is obtained for these two
positions, as we now discuss. The reactor neutrino flux decreases
rapidly with increasing baseline. So, a rather fair assumption is that
the flux is dominated by the sub-cluster defined by all reactors included in a radius $\lesssim 1000$ km. For the SURF and LNGS locations one finds that
these sub-clusters involve only 5 reactors with a 2.1 GW and 1.8 GW
average power, respectively.  For the Kamioka, SNOLAB and Boulby
locations, the sub-clusters are composed instead of 35, 59 and 49
reactors. The average power in each case (and in that order) is: 2.1
GW, 4.9 GW and 1.9 GW. Thus, already from these numbers one expects
the SURF and LNGS location sites to involve a less intense reactor neutrino flux.

Fig. \ref{fig:reactor_clusters} shows the distribution of nuclear
reactors in terms of baseline and power for the five different
clusters we consider. The distributions involve the full
data sets. From the graph, one can see that for the Boulby and SNOLAB
clusters the reactor density for baselines below 1000 km is high, with
a few of those reactors having powers above 3 GW. The distribution for
the Kamioka cluster is somewhat different. Although below 1000 km
there are a few reactors, their density is lower as well as their
power. For the SURF and LNGS clusters, the reactor density for
baselines below 1000 km is, instead, rather moderate. For these
clusters, most reactors are at baselines above 1000 km. So, even
without a dedicated calculation of the event rate, expectations are
that in terms of increasing reactor neutrino fluxes the clusters can
be sorted into three groups: SURF/LNGS, Kamioka, SNOLAB/Boulby.

The calculation of the differential nuclear recoil spectrum at each
cluster (C) requires the convolution of the differential CE$\nu$NS
cross section \cite{Freedman:1973yd,Drukier:1984vhf} with the reactor
neutrino flux, namely
\begin{equation}
  \label{eq:DnRS}
  \frac{dR_\text{C}}{dE_r}=
  \frac{m_\text{det}\,N_A\,\mathcal{T}\,\eta_\text{C}}
  {m^\text{Xe}_\text{mol}}
  \int_{E_\nu^\text{min}}^{E_\nu^\text{max}}
  \frac{d\Phi_{\overline\nu_e}}{dE_\nu}
  \frac{d\sigma}{dE_r}F^2_\text{H}(E_r)dE_\nu\ .
\end{equation}
Here, $m_\text{det}$ refers to the detector active volume mass,
$m_\text{mol}^\text{Xe}$ to the xenon molar mass, $\mathcal{T}$ to
the exposure time, $E_\nu^\text{min}=\sqrt{m_NE_r/2}$ ($E_r$ and $m_N$
refer to nuclear recoil energy and mass), and $E_\nu^\text{max}$ to the
neutrino spectrum kinematic ``high-energy'' tail taken at
$8\,$MeV.  
The average nuclear mass is
$\langle m_\text{Xe}\rangle/\text{GeV}=0.93\times \langle
A\rangle$, $\langle A\rangle=\sum_i X_iA_i=131.4$ being the mass number averaged over the nine stable xenon isotopes. We include---for completeness---the weak-charge nuclear
form factor, $F_\text{H}(E_r)$, parametrized \`a la Helm
\cite{Helm:1956zz}. Note that if not included results would deviate
from those presented here at most by $\sim 2\%$, because of the
process occurring deep in the full coherent regime.

Regarding the electron antineutrino spectrum, we proceed as follows. For the
$^{235}\text{U}$ and $^{238}\text{U}$ emission spectra we use results from
Ref.~\cite{Kopeikin:2021ugh}. For $^{239}\text{Pu}$ and
$^{241}\text{Pu}$ we use instead results from
Ref.~\cite{Huber:2011wv}. The full electron antineutrino differential flux is then calculated according
to
\begin{equation}
  \label{eq:anti_e_flux}
  \frac{d\Phi_{\overline\nu_e}}{dE_\nu}=\sum_{i=\text{Isotopes}}
  f_i\,\frac{d\Phi_{\overline\nu_e}^i}{dE_\nu}\ ,
\end{equation}
where
$f_i=\{f_{^{235}\text{U}},f_{^{238}\text{U}},f_{^{239}\text{Pu}},f_{^{241}\text{Pu}}\}=\{5.5,0.7,3.2,0.6\}\times
10^{-1}$ are the uranium and plutonium fission fractions
\cite{TEXONO:2006xds}. Note that we do not include electron
antineutrinos produced in neutron capture by $^{238}\text{U}$. The
reason is that the spectra for those neutrinos dominate at energies
below $\sim 1.5\,\text{~MeV}$, hence in a LXe detector would produce
nuclear recoils below 0.04 keV (much below any realistic operation
threshold). We assume the spectral function in
Eq.~(\ref{eq:anti_e_flux}) to be universal for all the reactors within
the clusters \footnote{Each reactor has its own fission fractions, but
  variations are at the permille level (see e.g. Tab.~4 in
  Ref.~\cite{Giunti:2021kab}).}. Thus, the difference among clusters
is determined only by the normalization factor, which we calculate
assuming that in each fission process an energy of
$\epsilon = 205.24$~MeV is released and that neutrinos are emitted
isotropically. Explicitly, each normalization factor is given by
\begin{equation}
\eta_\text{C} = \sum_{j} \frac{P_j}{4\pi L_j^2 \epsilon}\ ,
\end{equation}
where $j$ runs over all reactors relevant for cluster C and $P_j$ and
$L_j$ are the operating power and distance for reactor $j$. Their
values are displayed in Tab. \ref{tab:normalization}, showing that
SURF is subject to the least abundant neutrino flux, whereas Boulby to
the most severe.
\begin{table}[h]
  \centering
  \begin{tabular}{|c||c|c|c|c|c|}\hline
    Cluster& SURF & SNOLAB & Kamioka& LNGS& Boulby\\\hline
    $\eta_\text{C}$[$\text{cm}^{-2}\text{sec}^{-1}$]&
    20422&156630&103903&56677&932874\\\hline
  \end{tabular}
  \caption{Neutrino flux normalization factors for the five reactor
    clusters.}
  \label{tab:normalization}
\end{table}

\begin{figure*}[t]
  \centering
  \includegraphics[scale=0.43]{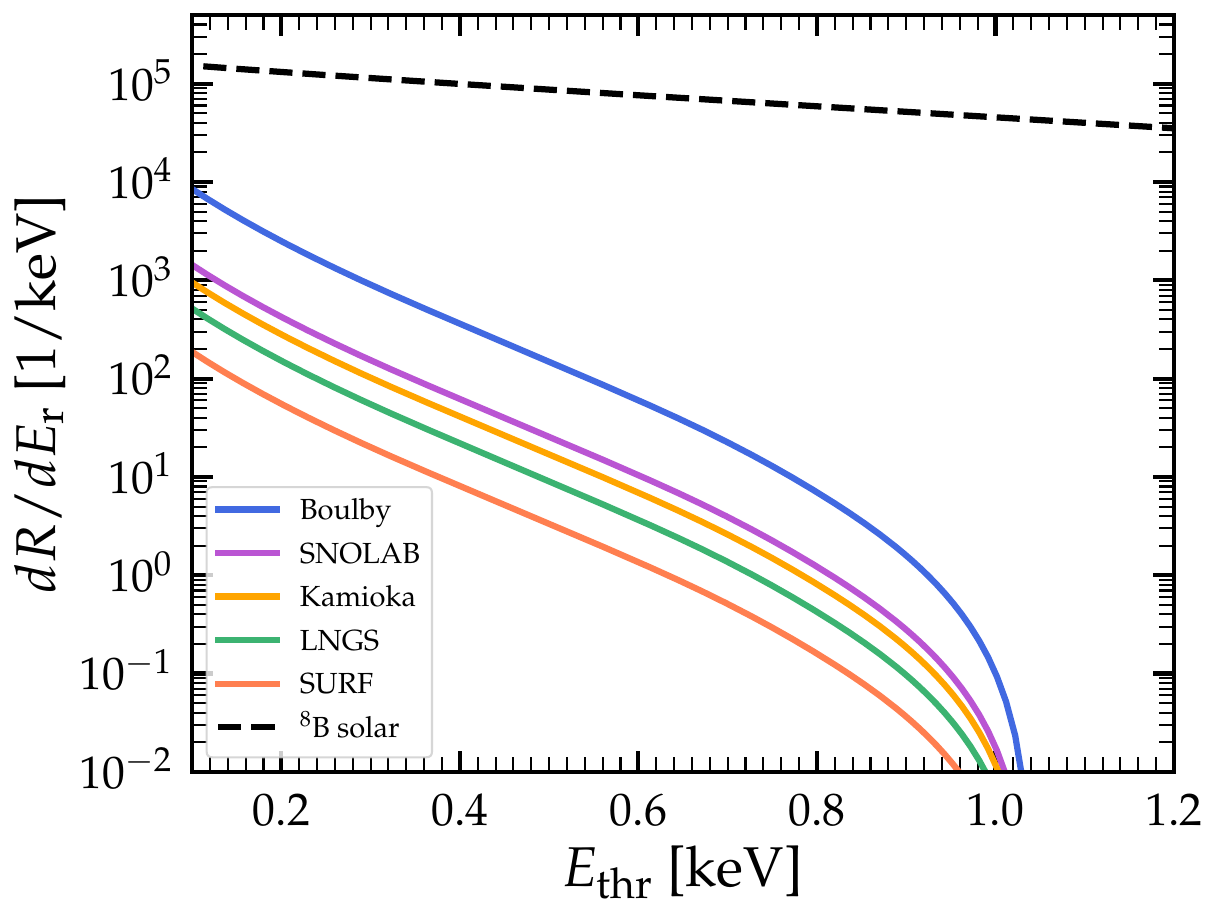}
  \includegraphics[scale=0.43]{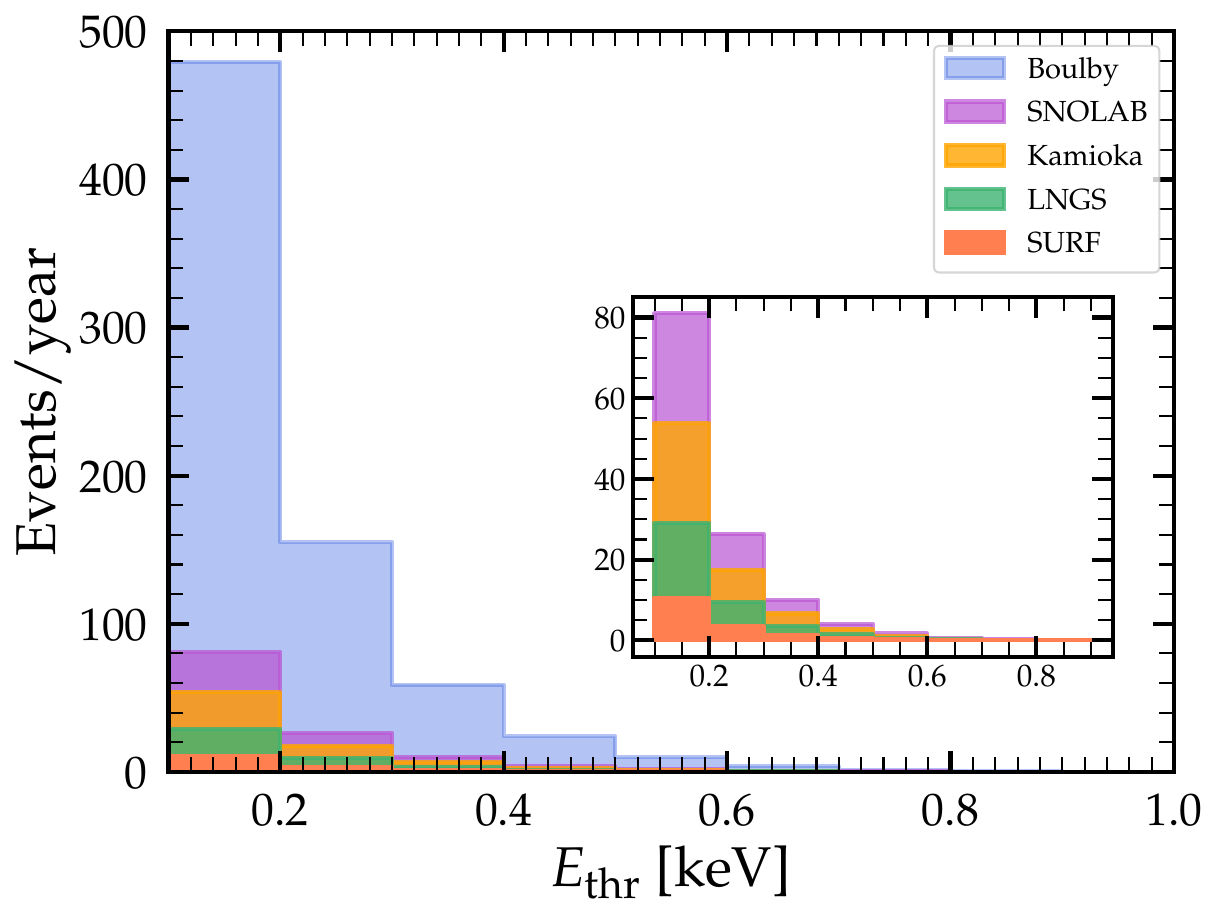}
  \caption{\textbf{Left graph}: Reactor neutrino differential event
    rate for the five detector sites considered in this work: SURF,
    SNOLAB, Kamioka, LNGS, and Boulby as a function of nuclear recoil
    energy. Shown as well is the $^8$B differential event
    rate. \textbf{Right graph}: Reactor neutrino total event rate for
    the same detector locations.}
  \label{fig:results}
\end{figure*}
With these results at hand, we are now in a position to calculate the
differential event rate as well as the total event rate for each
detector site. We assume a 50-tonne active volume LXe
detector and 100\% efficiency \footnote{The XLZD consortium aims at
  masses from 40 to 100 tonne. So this value is used just as a proxy
  of what the actual detector will use.}. Since current realistic
thresholds amount to 0.3 keV \cite{Lenardo:2019fcn}, we use
$E_r^\text{th,min}=0.1\,\text{keV}$ as a value envisioned for future
detector operations. Results are displayed in
Fig.~\ref{fig:results}. The left (right) graph shows the differential
event rate (total event rate) as a function of the recoil energy (recoil
energy threshold) for the five different reactor clusters we have
considered. The inset plot in the right panel is meant to zoom in on the bottom left corner.
Inline with expectations, the differential and total event
rates at the SURF (Boulby) detector site are the smallest
(largest). The event rate at the LNGS detector location is slightly
higher, followed by Kamioka and SNOLAB.
\section{Discussion}
\label{sec:discussion}
Naively one would expect the reactor neutrino flux to be suppressed and of little
relevance. This expectation is mainly based on the fact that most
reactors are far away from the detector sites. However, the fact that
the clusters around each detector site involve a large number of
active nuclear power plants (with in some cases powerful reactors),
combined with a large active volume produces a non-zero event rate in
all cases.

Ideally one would like a very low threshold to explore the small WIMP
mass window and increase the WIMP-nucleus event rate. At 0.1 keV, we
find that the total neutrino-nucleus event rate per year is: 16
(SURF), 44 (LNGS), 82 (Kamioka), 124 (SNOLAB) and 733 (Boulby). If
that operation threshold is not achieved and instead the detector is
operated at 0.3 keV, these numbers will be degraded by about a factor
7. In such an experimental scenario the reactor neutrino background
becomes, of course, less severe. Thus, the question of whether the
reactor neutrino background matters is---as anticipated---strongly
linked to operation thresholds.

It is worth emphasizing that variations of these estimated numbers are expected
in the future, depending on the exact number of reactors that enter in
either operation phase or are decommissioned. However, these results
demonstrate that the reactor neutrino flux should be seriously taken
into account in decision making as well as in data taken, contrary to
 expectations.

Finally, one might wonder how much this neutrino background matters compared
to the boron-8 solar neutrino flux. For the detector configurations
we have considered, with a 0.1 keV operation threshold, the number of
boron-8 nuclear recoil induced events is overwhelming, 36500
events/year. So, of course, this will be the dominant background
source. All the efforts to understand the morphology of this background are indeed motivated by this fact. The question is then
whether one should be concerned with the reactor neutrino background
whatsoever.

It is well known that the boron-8 background can be to a certain
degree circumvented. As we have already stressed, large data sets
might enable differentiating neutrino from WIMP signals, if the WIMP
parameters are such that the neutrino and WIMP event rates strongly
degenerate. In general, however, directional detectors seem to be the
most promising avenue \cite {Vahsen:2020pzb,Vahsen:2021gnb}
\footnote{They have been as well recently considered for CE$\nu$NS
  measurements and beyond the SM searches using neutrino beamlines at
  Fermilab
  \cite{Abdullah:2020iiv,AristizabalSierra:2021uob,AristizabalSierra:2022jgg}}. For these detectors it seems that the reactor neutrino background might even become the most dominant background source. Therefore, if the boron-8 nuclear recoil-induced events
can be efficiently discriminated, there will be yet another background
source that will require careful identification and proper treatment, depending on statistics and operation capabilities.

\section{Conclusions}
\label{sec:conclusions}
With the advent of third-generation DM direct detection detectors, the quantification of
reactor neutrino fluxes becomes of pivotal importance.  In this work
we have quantified the size of the neutrino flux produced by clusters
of reactors surrounding five potential detector deployment sites. For
definitiveness we have considered the locations envisioned by the
recently established XLZD consortium: SURF, SNOLAB, Kamioka, LNGS, and
Boulby.

Our findings show that detectors with active volumes of the order of
50 tonne and recoil energy thresholds of the order of 0.1 keV, will be
sensitive to a certain amount of reactor neutrino-induced events.  
The exact amount depends, to a large degree, on the energy threshold at which the
detector is operated. However, even assuming a realistic 
threshold of 0.3 keV, the event rate turns out to be sizable in all cases. We find that the
site with the smallest reactor neutrino background is SURF followed by LNGS, Kamioka, SNOLAB, and Boulby (in that order).

Although subdominant compared to the solar boron-8 neutrino background, we
point out that the reactor neutrino background (and its corresponding
events) should be---in principle---considered during data
taken. Reactor neutrino-induced events should be taken into account in
background discrimination, regardless of the detector technique
employed. This result will be particularly relevant for directional detection, if future
detectors meet the requirements we have used here.
\section*{Acknowledgments}
We thank P. Martínez-Miravé for pointing out to us the
Geoneutrinos.org website.  The work of D.A.S. is funded by ANID under
grant ``Fondecyt Regular'' 1221445. He thanks ``Le Service de Physique
Th\'eorique (Universit\'e Libre de Bruxelles)'' and ``Instituto de
F\'{\i}sica Corpuscular (CSIC y Universidad de Valencia)'' for their
kind hospitality and their stimulating research environment during the
completion of this work.
V.D.R. acknowledges financial support from the CIDEXG/2022/20 grant
(project ``D'AMAGAT'') funded by Generalitat Valenciana and by the
Spanish grant PID2020-113775GB-I00 (MCIN/AEI/10.13039/501100011033).
C.A.T. is very thankful for the hospitality at Universidad T\'ecnica
Federico Santa Mar\'{i}a, where this work was initiated.
\bibliographystyle{utphys}
\bibliography{references}
\end{document}